\documentclass[10pt, conference]{IEEEtran}
\IEEEoverridecommandlockouts
\usepackage{graphicx} 
\usepackage{xspace}
\usepackage{pgfplots}
\usepackage{pifont}
\usepackage{multirow}
\usepackage{balance}
\pdfoutput=1
\usepackage[T1]{fontenc}
\usepackage[utf8]{inputenc}
\usepackage{newtxtext}
\usepackage{newtxmath}

\usepackage{textcomp}
\usepackage{bm} 

\usepackage{tcolorbox}
\pgfplotsset{compat=1.18}
\usepgfplotslibrary{groupplots}
\usepackage{soul}
\usepackage{hyperref}
\usepackage{multirow}
\usepackage{listings}
\usepackage{fancybox}
\usepackage{enumitem}
\usepackage{graphicx}
\usepackage{color}
\usepackage{array}
\usepackage{amsmath}
\usepackage{xspace}
\usepackage{url}
\usepackage{tikz}
\usepackage{caption}
\usepackage{pgfplots}
\usepackage{balance}
\usepackage{subcaption}
\pgfplotsset{compat=1.16}
\usepackage{pgf-pie}
\usetikzlibrary{pgfplots.statistics,calc}
\usepackage{algorithm}
\usepackage{algorithmic}
\usepackage{adjustbox}
\usepackage{makecell}

\usepackage{tcolorbox}
\makeatletter
\newcommand{\mybox}[1]{%
	\setbox0=\hbox{#1}%
	\setlength{\@tempdima}{\dimexpr\wd0+13pt}%
	\begin{tcolorbox}[boxrule=0.5pt, colback=white, arc=4pt,
		left=6pt,right=6pt,top=6pt,bottom=6pt,boxsep=0pt]
		#1
	\end{tcolorbox}
}

\definecolor{songcolor}{RGB}{191,191,191}

\newcommand*\circled[1]{\tikz[baseline=(char.base)]{
            \node[shape=circle,draw,inner sep=2pt] (char) {#1};}}

\newcommand{\tool}{SPARK~}

\usepackage{colortbl}  
\usepackage{xcolor}    

\definecolor{lightblue}{RGB}{220,230,241}
\definecolor{mygreen}{RGB}{0,150,0}
\definecolor{myred}{RGB}{200,0,0}
\definecolor{mygray}{RGB}{128,128,128}

\usepackage{xcolor, listings, booktabs}
\usepackage{subcaption}
\usepackage{hyperref}

\lstdefinestyle{pymini}{
  language=Python, basicstyle=\ttfamily\footnotesize, numbers=left, numberstyle=\tiny,
  showstringspaces=false, frame=single, framerule=0.4pt, columns=fullflexible
}
\def\BibTeX{{\rm B\kern-.05em{\sc i\kern-.025em b}\kern-.08em
    T\kern-.1667em\lower.7ex\hbox{E}\kern-.125emX}}

\begin{document}

\title{\tool: \underline{S}emantic-preserving \underline{P}rompt Mutation for Code Gener\underline{a}tion using Consensus-based Ve\underline{r}ification Framewor\underline{k}\thanks{This work
was supported by the NSERC Discovery Grant and Vector Institute Scholarship in AI.
}}
\newif\ifanonymous
\anonymousfalse

\ifanonymous
\author{
    \IEEEauthorblockN{Anonymous Author(s)}
    \IEEEauthorblockA{Author information withheld for double-blind review.}
}
\else
\author{
    \IEEEauthorblockN{Alireza Daghigh}
    \IEEEauthorblockA{\textit{York University}\\
    Toronto, Canada\\
    alirezadeq@yorku.ca}
    \and
    \IEEEauthorblockN{Chung-Yu Wang}
    \IEEEauthorblockA{\textit{York University}\\
    Toronto, Canada\\
    cywang14@yorku.ca}
    \and
    \IEEEauthorblockN{Luca Filippelli}
    \IEEEauthorblockA{\textit{York University}\\
    Toronto, Canada\\
    lucafili@my.yorku.ca}
    \and
    \IEEEauthorblockN{Hamed Taherkhani}
    \IEEEauthorblockA{\textit{York University}\\
    Toronto, Canada\\
    hamedth@yorku.ca}
    \and
    \IEEEauthorblockN{Hadi Hemmati}
    \IEEEauthorblockA{\textit{York University}\\
    Toronto, Canada\\
    hemmati@yorku.ca}
    \and
    \IEEEauthorblockN{Hung Viet Pham}
    \IEEEauthorblockA{\textit{York University}\\
    Toronto, Canada\\
    hvpham@yorku.ca}
}
\fi

\date{January 2025}


\maketitle
\begin{abstract}
Automated code generation with large language models (LLMs) has improved in recent years. With the lack of test cases in real-world usages, generated code is not automatically verifiable and often requires manual validation. Previous studies have leveraged consensus verification, where code solutions and corresponding test cases are generated and evaluated based on functional consistency. Building on this, SPARK incorporates semantic-preserving mutations to problem descriptions such as paraphrasing, grammatical adjustments, function renaming, and other alterations to enhance the robustness and correctness of LLM-generated code. For each mutated description, SPARK generates a set of code solutions and test cases. These candidate solutions are evaluated using consensus-based verification, where the solutions' quality is measured by how consistently they pass generated tests. We assess the effectiveness of SPARK on three benchmarks (LiveCodeBench, BigCodeBench, and DLBench) across three LLMs (o4-mini, DeepSeek-R1-Basic, and Gemini-2.5-Pro). Our evaluation demonstrates that SPARK outperforms state-of-the-art consensus-based verification approaches up to 4.1\%. Our generalization experiment highlights the practical potential of SPARK as an effective strategy for enhancing the reliability and correctness of LLM-based code generation systems in scenarios where traditional validation resources are scarce, while our qualitative analysis shows  insight into how SPARK’s mutation helps LLMs generate better code solutions.

\end{abstract}


\section{Introduction}
Automated code generation~\cite{odeh2024comparative, shin2021survey} with large language models (LLMs)~\cite{jiang2024survey, chen2023duetcs, hou2024large, fan2023large}
has advanced rapidly in recent years, offering promising tools for software development. Despite these gains, reliably producing a single correct, executable solution remains a significant challenge~\cite{becker2023programming, odeh2024comparative, ridnik2024code}. Many models achieve high success rates when allowed to generate multiple candidate solutions, as measured by metrics such as \emph{pass@k} (the probability that at least one of the top-\emph{k} generated solutions passes the test suite)~\cite{lyu2025top, liu2023your}. However, their top-ranked outputs often fail to meet correctness requirements in practice~\cite{nijkamp2022codegen}. This gap highlights the practical difficulty of selecting a reliable solution from multiple generated candidates, especially in real-world scenarios where comprehensive, human-written test suites are scarce or unavailable~\cite{shin2024retrieval, wang2024testeval, xu2011tool}.

\begin{figure}
    \centering
    \includegraphics[width=0.9\linewidth]{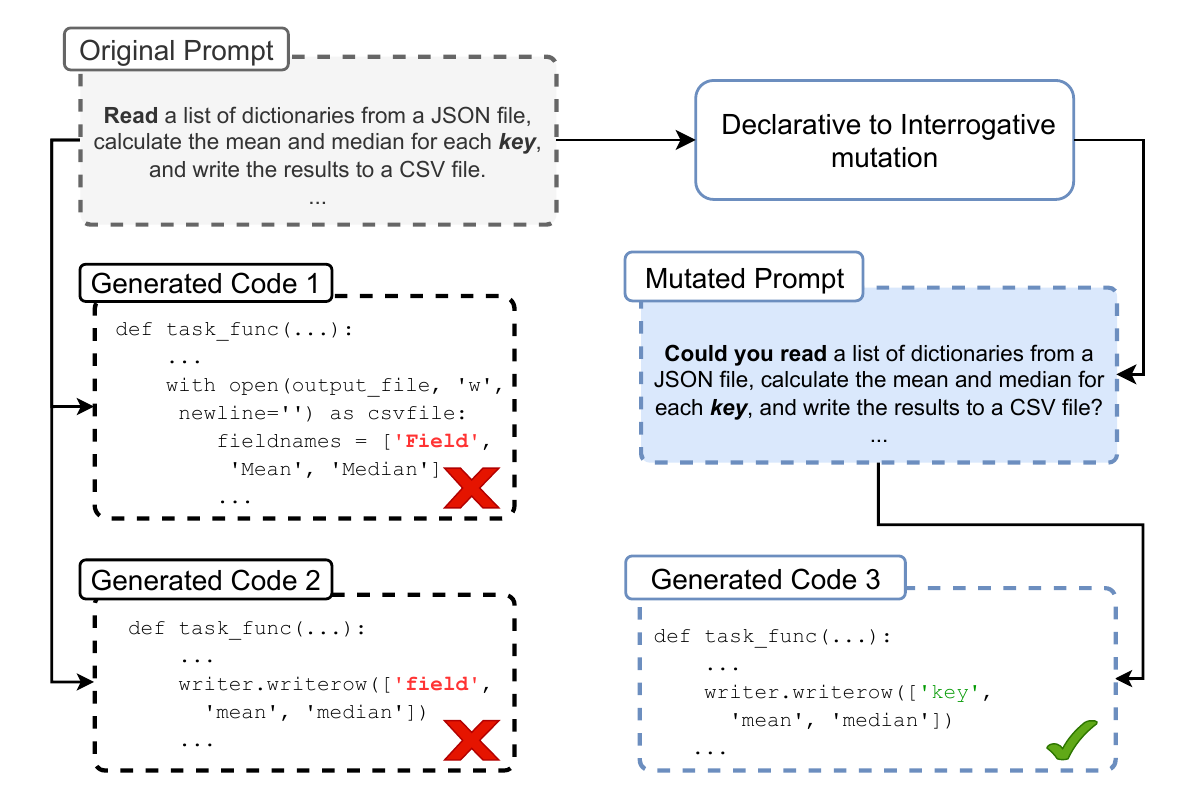}
    \caption{
    An example showing a prompt mutation from declarative to interrogative leads the LLM to generate code with a corrected CSV header (\texttt{key} instead of \texttt{field}).
    }
    \label{fig:dual_success}
    \vspace{-7pt}
\end{figure}
To address the lack of ground-truth tests in practice, prior work such as the CodeT framework~\cite{chen2022codet} leverages generated corresponding test cases to verify code solutions. Inspired by consensus-based algorithms~\cite{bonandin2024consensus} such as RANSAC~\cite{derpanis2010overview}, this \textit{dual agreement} approach can identify groups of solutions that consistently pass groups of generated test cases and thus exhibit highly cohesive functionality.
CodeT demonstrated improvements in code generation benchmarks without relying on hand-crafted test cases. However, since CodeT relies on a single prompt, the generated solutions are often tied to a single 
LLM's understanding of that one specification, leading to some misinterpretation. Figure~\ref{fig:dual_success} illustrates such a failure: when tasked with computing the mean and median from a JSON file and writing the results to a CSV, the original prompt repeatedly fails validation because the solution outputs a \texttt{field} header while the validation suite expects a \texttt{key} header. This scenario demonstrates how single-prompt generation can fail due to subtle specification ambiguities.

To mitigate this shortcoming, we propose \tool, which introduces \emph{semantic-preserving prompt mutations} to systematically create diverse, semantically equivalent variants of the original problem description. These instances increase diversity in both generated solutions and tests, improving the chances that more candidate matches the intended specification. Concretely, \tool  uses a common, language-focused LLM (GPT-4o) to apply controlled changes in phrasing, grammatical structure, function naming, and other lexical/syntactic features (see Table~\ref{tab:mutations}). For each mutated prompt, we generate code solutions and test cases; all candidates then participate in a dual-execution agreement selection. Candidate solutions are then grouped by their test-case pass patterns under the hypothesis that correct implementations exhibit consistent functional behavior across differently phrased prompts. We then select the cluster with the strongest agreement as the final output. As shown in Figure~\ref{fig:dual_success},  reformulating the original declarative instruction as an interrogative prompt leads the model to produce the correct \texttt{key} header, enabling it to pass the validation suite.

We evaluate \tool on three complementary benchmarks, LiveCodeBench~\cite{jain2024livecodebench} (algorithmic/data-structure tasks), BigCodeBench~\cite{zhuo2024bigcodebench} (real-world repository/API tasks), and DLBench~\cite{daghighfarsoodeh2025dl} (deep learning–oriented code generation) using three state-of-the-art LLMs: O4-Mini~\cite{jurisechatgpt}, DeepSeek R1 Basic~\cite{deng2025exploring}, and Gemini 2.5 Pro~\cite{team2023gemini}. All prompt mutations are produced with GPT-4o. 

Our experiments show that incorporating a single prompt mutation in most cases already improves the \emph{pass@1} performance across most dataset–model combinations. While the magnitude of improvement varies, we observe clear patterns. For example, verb-based modifications (e.g., converting between active and passive forms or replacing verbs with near-synonyms) provide the strongest and most reliable gains, particularly on LiveCodeBench and DLBench.
Context-based modifications, such as altering or removing function names, also contribute meaningfully, particularly on repository-style tasks in BigCodeBench. 
Styling-oriented mutations, though more superficial, still produce measurable improvements by prompting models toward more precise code generation. 
These results suggest that different classes of mutations address complementary weaknesses in LLMs, with verb modifications enhancing semantic alignment, context modifications improving robustness to naming and specification noise, and styling changes introducing additional syntactic diversity.

We expand upon the analysis of individual mutations by also studying the \emph{compositions} of mutations. We first explore combinations of mutations on a calibration dataset to identify the best-performing mutation set. Such combinations are then applied to the tasks from a different dataset. Specifically, on average, \tool achieves up to 4.1\% improvement in pass@1 score over CodeT, the state-of-the-art consensus-based validation technique. This experiment demonstrates \tool's expandability and practical usage in a real-world scenario where additional mutation can be added, and tuning on the calibration dataset is generalizable to deployment time.

Our qualitative analysis shows that mutations improve code generation through several mechanisms. The most common is generating diverse implementations that reduce overfitting to a single phrasing. Mutations also increase resilience to misleading or distracting context, clarify implicit requirements, expose boundary and corner cases, and act as a filter against overfitted solutions by aligning with harder tests. Together, these effects explain why mutations often lead to stronger and more reliable programs. 

Overall, \tool not only raises accuracy in diverse benchmarks but also demonstrates that systematic mutation strategies can be tuned to the strengths of specific models and datasets, offering a practical path to more dependable LLM-based code generation. This paper makes the following contributions:
\begin{itemize}[leftmargin=*]
    \item We propose \tool, a code generation technique that combines semantic-preserving prompt mutations with consensus-based verification to select reliable code solutions.
    \item We conduct an extensive evaluation with three LLMs on three complementary code-generation benchmarks, showing consistent improvements over baselines of up to 4.1\% pass@1. 
    \item We provide qualitative insights into when and why different mutation strategies help or hurt, and we study a general strategy to select effective mutation \emph{combinations} that can generalize in real-world scenarios.
\end{itemize}

\noindent\textbf{Data Availability: }
We release our code and data\footnote{https://github.com/Smart-System-testing-lab/SPARK}.
\section{Approach} \label{approach}

\begin{figure*}
    \centering
    \includegraphics[width=0.9\textwidth]{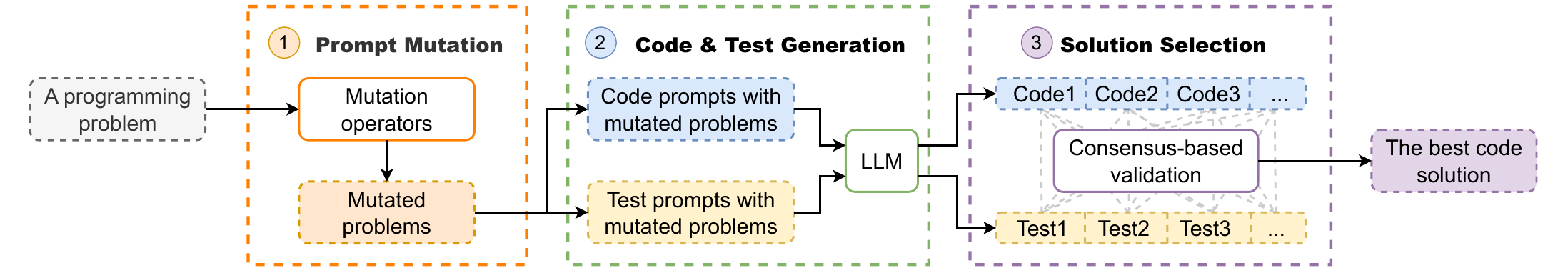}
    \caption{
    Overview of \tool
    }
    \label{fig:pipeline}
    \vspace{-7pt}
\end{figure*}

Figure~\ref{fig:pipeline} provides \tool's overview. When given \textbf{\textit{a programming problem}}, \tool first performs \circled{1} \textbf{Prompt Mutation}. 
Using the original and mutated problems, \tool then queries the LLM to execute the \circled{2} \textbf{Code \& Test Generation} step. 
By executing the generated test cases and applying consensus-based validation, \tool identifies \textbf{\textit{the best code solution}} in the \circled{3} \textbf{Solution Selection} step. 

\begin{table*}[ht]
\centering
\renewcommand{\arraystretch}{1.5}
\caption{
Mutation operation, their class,  their prompt template, and corresponding input–output problem descriptions.
}
\label{tab:mutations}
\resizebox{0.9\linewidth}{!}{%
\begin{tabular}{@{}lllll@{}}
\toprule
\large
\textbf{Class} & \textbf{Operation (Abbr.)} & \textbf{Mutation Prompt Template} & \textbf{Example Input}  & \textbf{Example Output} \\ 
\midrule

\multirow{4}{*}{\makecell[l]{Context \\Modification}} 
& Remove Function Name (RFN) & 
\makecell[l]{Modify the following sentence to exclude references \\ to the function's name} & 
\makecell[l]{Generate a Python function \\ \textbf{\textit{multinomial\_sampling}}...} & 
Generate a Python function... \\ 

& Unrelated Function Name (UFN) &
\makecell[l]{Modify the following sentence to make the function's \\ name unrelated to its meaning}& 
\makecell[l]{Generate a Python function \\ \textbf{\textit{multinomial\_sampling}}...} & 
\makecell[l]{Generate a Python function \\ \textbf{\textit{chocolate\_cheesecake}}...}\\ 

& Adversarial Function Name (AFN)& 
\makecell[l]{Modify the following sentence to refer to a function \\ with a misleading name that doesn't represent the \\ actual task} & 
\makecell[l]{Generate a Python function called \\ \textbf{\textit{get\_perspective\_transform}}...}& 
\makecell[l]{Generate a Python function called \\ \textbf{\textit{compute\_angle\_of\_projection}}...}  \\ 

& Rename Variables (RV) & 
\makecell[l]{Modify the following sentence to rename any \\ explicitly named variable} &
\makecell[l]{Generate a Python function which \\ takes an attribute named \textbf{\textit{var1}}...} &
\makecell[l]{Generate a Python function which \\ takes an attribute named \textbf{\textit{var2}}...} \\ 

\midrule

\multirow{4}{*}{\makecell[l]{Styling \\Modification}}
& Lowercase to Uppercase (L2U)& 
Change the following sentence to uppercase letters& 
\makecell[l]{G\textbf{\textit{enerate a }}P\textbf{\textit{ython function called}} \\ \textbf{\textit{hinton}}...} &
\makecell[l]{G\textbf{\textit{ENERATE A }}P\textbf{\textit{YTHON}} \\ \textbf{\textit{FUNCTION CALLED HINTON}}... }\\

& Uppercase to Lowercase (U2L)&
Change the following sentence to lowercase letters  & 
\makecell[l]{\textbf{\textit{G}}enerate a \textbf{\textit{P}}ython function called \\ hinton...} &
\makecell[l]{\textbf{\textit{g}}enerate a \textbf{\textit{p}}ython function called \\ hinton...}\\

& Star Highlight (SH)& 
\makecell[l]{Encase important phrases with two asterisks per \\ side} &
\makecell[l]{Generate a \textbf{\textit{Python function}} called \\ hinton...} &
\makecell[l]{Generate a \textbf{\textit{**Python function**}} \\ called hinton...} \\

& Uppercase Highlight (UH)&
Highlight important phrases with uppercase letters &
\makecell[l]{Generate a Python function called \\ hinton...} &
\makecell[l]{Generate a \textbf{\textit{PYTHON FUNCTION}} \\ called hinton... }\\

\midrule

\multirow{5}{*}{\makecell[l]{Verb \\Modification}}
&  Active to Passive (A2P)& 
Change the following sentence to the passive voice & 
\makecell[l]{\textbf{\textit{Write}} a Python function \\ cot\_laplacian that computes...}&
\makecell[l]{A Python function cot\_laplacian \\ \textbf{\textit{should be written to}} compute...} \\

& Passive to Active (P2A)&
Change the following sentence to the active voice &
\makecell[l]{A Python function  cot\_laplacian \\ \textbf{\textit{should be written to}} compute... }&
\makecell[l]{\textbf{\textit{Write}} a Python function cot\_laplacian \\ that computes...} \\

&Present To Future Tense (2F)&
Change the following text to future tense phrasing &
\makecell[l]{\textbf{\textit{Write}} a Python function for the \\ following task... }& 
\makecell[l]{A Python function \textbf{\textit{will be written}} \\ for the following task:... }\\

&Present To Past Tense (2P)&
Change the following text to past tense phrasing &
\makecell[l]{\textbf{\textit{Write}} a Python function for the \\ following task... }& 
\makecell[l]{A Python function \textbf{\textit{was written}} \\ for the following task:... }\\

 & Verb to Similar Verb (V2V)&
 \makecell[l]{Change the verbs of the following sentence to a \\ similar verb } &
 \makecell[l]{\textbf{\textit{Write}} a Python function for the \\ \textbf{\textit{following}} task: \textbf{\textit{Describe}}... } &
 \makecell[l]{\textbf{\textit{Construct}} a Python function for \\ the \textbf{\textit{subsequent}} task: \textbf{\textit{Illustrate}}... } \\

\midrule
 
\makecell[l]{Tone \\Modification }& 
Declarative to Interrogative (D2I)& 
\makecell[l]{Change the following sentence to interrogative \\ voice }& 
\makecell[l]{\textbf{\textit{Generate}} a Python function called \\ top\_k\_accuracy... }&
\makecell[l]{\textbf{\textit{Is it possible to generate}} a Python \\ function  called  top\_k\_accuracy... }\\  
 
\bottomrule
\end{tabular}
}
\vspace{-7pt}
\end{table*}
\subsection{Prompt Mutation}
One of the strengths of the consensus-based validation process proposed by CodeT is that it can estimate the solution quality by verifying the majority agreement of code solutions with generated test cases. However, in cases where the LLM failed to generate correct code solutions from the original prompt, CodeT would select the incorrect solutions. \tool utilizes a set of semantic-preserving mutations of the problem prompt to help trigger additional correct solutions.
As the example in Figure~\ref{fig:dual_success} demonstrates, 
a tone modification from declarative to interrogative
can induce more diversity in the generated code, thereby helping the LLM produce additional correct solutions that can improve the overall accuracy of the code generation system.

In step \circled{1}, \tool influences solution diversity by creating problem description variance through semantic-preserving mutation operations. These operators aim to vary the prompt styling, tone, syntactic structure, or context. This form of variation can trigger changes in the generated code, which can improve the overall correctness, efficiency, and readability of the candidate solution set (refer to \ref{tab:mutation_discussion} for further discussion).
Specifically, \tool applies fourteen \textbf{prompt mutations} operators to the original problem description in four major thematic classes: Context Modification, Styling Modification, Verb Modification, and Tone Modification. 

Table~\ref{tab:mutations} shows the mutation prompt template and an example for each mutation operation. Each mutation class aims to introduce changes that can induce additional solution generation:

\begin{itemize}[leftmargin=*]
    \item \textbf{Context Modification} focuses on mutating the context of the code generation task. For example, some problem descriptions include function names that can influence the model to generate a specific solution. By mutating this context, the prompt variances can influence different solutions to be generated.
    \item \textbf{Styling Modification} focuses on modifying the styling and highlighting of the problem descriptions. Different highlighting and styling can influence the attention of the models, which can induce additional solution variance.
    \item \textbf{Verb Modification} focuses on modifying the verbs tenses. Modifying these action verbs can directly induce changes in the generated solutions.
    \item \textbf{Tone Modification} focuses on the tone of the request. Changing the tone could indirectly instigate additional solution variances.
\end{itemize}

\subsection{Code and Test Generation}  
Once the prompt variances are collected, \tool queries LLMs to generate both code solutions and test cases to prepare for the consensus-based validation. In step \circled{2}, \tool utilizes different prompt templates for code and test generation for each problem description to generate candidate code solutions and test cases.

\noindent\textbf{ Code Solution Generation: }
To generate code solutions, \tool directly queries LLMs with the problem description. Since the origin and mutated problem description already contain code generation commands, no additional request is needed. Due to the generated code sometimes containing missing imports, inconsistent function names compared to those expected in the test cases, or mismatched variable names between the function definition and the test inputs, additional post-processing is needed. This includes automatically inserting or correcting import statements, renaming functions to align with the required signatures, and adjusting variable names so that arguments match the test cases. Each generated code solution is then added to the solution candidate pool.

\noindent\textbf{ Test Case Generation: }
In addition to generating code solutions, \tool also generate test cases which will be used in the consensus-based validation phase to estimate the quality of solution candidates.
Specifically, a \textbf{test case} includes a specific input along with the expected output for the function defined in the problem context. For example, in a task that determines whether there are elements in a list that are closer than a given threshold, a test case would provide an input list and threshold value, as well as the correct Boolean result.  


\begin{figure*}[h!]
  \centering
  \subfloat[Input-output assertion\label{fig:input_output}]{
    \includegraphics[width=0.20\linewidth, trim=0 0 0 0cm, clip]{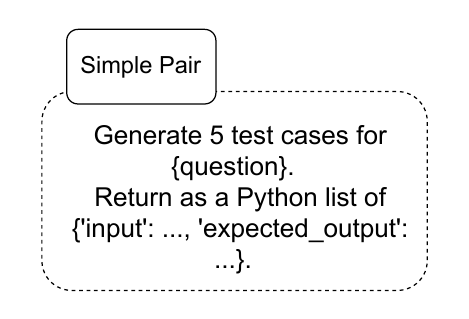}
  }
  \hfill
  \subfloat[Unittest format\label{fig:unittest}]{
    \includegraphics[width=0.20\linewidth, trim=0 0 0 0cm, clip]{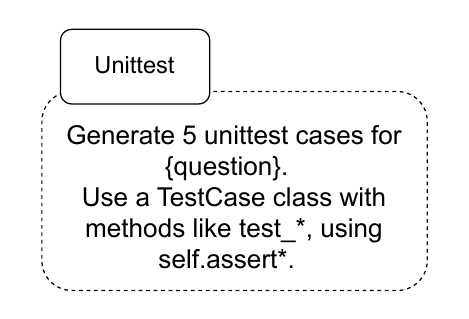}
  }
  \hfill
  \subfloat[Pytest format\label{fig:pytest}]{
    \includegraphics[width=0.209\linewidth, trim=0 0 0 0cm, clip]{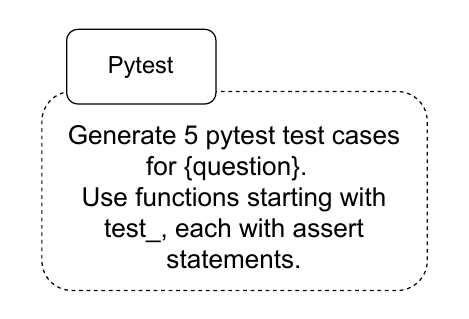}
  }
  \caption{Prompt templates for test case generation.}
  \label{fig:test_template}
  \vspace{-7pt}
\end{figure*}

To generate these test cases, instead of directly querying LLMs using the original problem description, we apply a test generation template (Figure~\ref{fig:test_template}). This template replaces the code generation command with the test case generation command.
Similar to code generation, \tool generates test cases with both the original problem description and the mutated ones and adds them to the test cases pool.

Specifically, our template
tailors the test case generation to support different benchmark requirements. Specifically, we produce three distinct types of test cases to match the expected formats of the major usages corresponding to the major benchmarks:

\begin{itemize}[leftmargin=*]
    \item \textbf{Simple input-output assertions}: These are direct assert statements which check that the function produces the correct output for a given input. This style is compatible with the \textit{LiveCodeBench} benchmark, which requires concise, clear input-output pairs.
    
    \item \textbf{Unittest format}: These test cases follow the structure of Python's \texttt{unittest} framework, organizing checks into test classes and methods. This format aligns with the needs of \textit{BigCodeBench}, which uses standard unit testing conventions.
    
    \item \textbf{Pytest format}: These tests use the \texttt{pytest} style, often employing function-based tests with decorators or simple assertions that integrate easily with the \texttt{pytest} runner. This format ensures compatibility with \textit{DL-Bench}, which adopts \texttt{pytest} as its testing standard.
\end{itemize}

In terms of testing frameworks, our method considers the two most popular libraries in Python, along with the default assertion style. By capturing these testing formats, our approach generalizes to most development environments. In the event the above test styles are not used, the user can create a new template by making trivial modifications to the existing prompt templates. 
By generating test cases in these varied formats, we evaluate the generalization of our approach to any development environment with various unique testing requirements.

\noindent\textbf{Solutions Selection:}
In step \circled{3}, \tool executes the test cases to validate the code solutions in a consensus-based validation process (i.e., dual agreement\cite{chen2022codet}). Since the generated test cases are not guaranteed to be correct, this step estimates the quality of the solution candidates. While we directly adopt the original validation step, our diverse prompts help tease out more correct solutions, hence overall improve the validation process effectiveness.

\noindent\textbf{\textit{Tests execution: }}
To compute the quality score for each code solution candidate, \tool first validates each solution candidate by running it against all test cases in the pool.  To execute the test cases, \tool needs to perform different execution steps for different development environments:

\begin{itemize}[leftmargin=*]
    \item For \textbf{simple input--output assertion} tests, \tool wraps the candidate in a lightweight runner that (i) feeds the provided stdin to the program or function, (ii) captures stdout and stderr, (iii) normalizes whitespace and line endings, and (iv) compares the observed output against the expected string(s). Each case is executed in a fresh process with a wall-clock timeout and resource limits; non-zero exit codes, timeouts, or mismatches are recorded as failures.
    
    \item For \textbf{pytest} tests, \tool materializes a temporary project (candidate file plus tests), ensures the target symbol is importable (fixing the module path and function name if needed), and runs \texttt{pytest -q} with per-test timeouts. It captures pass/fail/skip/xfail outcomes and assertion messages.
    
    \item For \textbf{unittest} tests, \tool discovers and runs the suite via \texttt{unittest} loaders, importing the candidate module, applying any required shim for function renaming, and executing each \texttt{TestCase} with buffered output and timeouts. It collects per-test outcomes (OK/FAIL/ERROR/SKIP) and condensed tracebacks.
\end{itemize}



\noindent\textbf{\textit{Consensus-based validation:}}
Once the pass and fail statistics are collected, \tool applies consensus-based validation by
computing a score for each consensus set. Following prior work~\cite{chen2022codet}, the score
is defined as 
\begin{equation}
f(S) = |S_x| \cdot |S_y| ,
\label{eq:dual-agreement}
\end{equation}
where:
\begin{itemize}[leftmargin=*]
    \item \(x \in X = \{x_1, x_2, \ldots, x_N\}\) denotes a \emph{code solution} generated by the model,
    \item \(y \in Y = \{y_1, y_2, \ldots, y_M\}\) denotes a \emph{test case} generated for the problem,
    \item \(S_x \subseteq X\) is the set of code solutions that pass exactly the same subset of test cases,
    \item \(S_y \subseteq Y\) is the set of test cases passed by all code solutions in \(S_x\).
\end{itemize}
Intuitively, the score measures the strength of a consensus set by combining the number of
agreeing solutions (\(|S_x|\)) and the number of test cases they jointly satisfy (\(|S_y|\)).
A higher score indicates stronger evidence that the corresponding functionality is correct.


Once the scores are computed. \tool ranks and selects the highest-scoring solutions. However, the consensus-based nature of the dual execution agreement may result in multiple candidate solutions passing the same set of tests, which results in multiple candidates with the same score. In such cases, we introduce a \textbf{tiebreaker mechanism} based on comparing the average scores of the tied solutions' code quality metrics. 
This approach assumes that the solution with higher structural quality is more likely to generalize and align with the ground-truth reference when functional behaviour is indistinguishable. 
To quantify code quality, we extract established metrics from the previous work~\cite{pantiuchina2018improving,maintability_index}: (i) the Maintainability Index (MI), which aggregates readability and complexity on a 0–100 scale (higher is better); (ii) Lack of Cohesion (LCOM1), where lower values indicate stronger cohesion among functions; (iii) Coupling Between Objects (CBO), which captures the number of distinct external modules used (lower is better); and (iv) Cyclomatic Complexity (CC), measuring the decision complexity of functions (lower is better). These metrics are combined into a single normalized score to represent the overall structural quality of each program.

This secondary ranking allows \tool to prefer programs that are not only correct on the given tests but are also more maintainable and less error-prone in practice. Code solutions that are ranked highest in both rankings are selected as \tool's final solutions.

\section{Experimental Setup}

\noindent\textbf{ Large Language Models}
Since semantic-preserving prompt mutation is a natural language task that is LLM's strength, in general, any recent LLM is sufficient. Hence, \tool employs GPT-4o, a low-cost general-purpose LLM that is capable of such tasks~\cite {lee2025metamon, li2024mutation}.
On the other hand, code and test generation is a complex task that requires much more sophisticated LLMs. We evaluate \tool on three state-of-the-art code generation models (o4-mini, DeepSeek R1 Basic, and Gemini 2.5 Pro) to ensure a robust comparison across different architectures and design philosophies. 
For fair evaluation, all models are queried in zero-shot mode using identical decoding parameters, with a temperature of 0.3, and top-p of 1.0.

\noindent\textbf{ Code Generation Benchmarks}
We evaluate \tool on three complementary benchmarks with varied developmental environmental settings and complexity. \textbf{LiveCodeBench} includes a suite of algorithmic and data-structure problems drawn from community coding challenges, testing general problem-solving ability. \textbf{BigCodeBench} consists of large-scale code synthesis tasks extracted from real-world repositories, emphasizing API usage, project structure, and realistic development practices. Finally, \textbf{DLBench} is a deep-learning–specific benchmark that covers model definitions, training loops, and tensor manipulation tasks, providing a targeted evaluation for machine learning workflows. These diverse benchmarks create a complementary evaluation experiment that assesses \tool's generalization and effectiveness across a broad range of programming tasks.

\noindent\textbf{ Evaluation Metric}
Following common practice, we adopt the \textbf{pass@k} metric, which measures the probability that at least one of the top-\(k\) generated solutions correctly solves the given problem, meaning 
it can pass all the ground-truth test cases.

\section{Result} 
\label{result}
In this section, we present the findings to answer our research questions that aim to investigate the generalization and effectiveness of \tool in code generation on three complementary benchmarks.



\subsection{RQ1: How well do LLMs perform prompt mutation while preserving problem description semantics?} 


This research question aims to assess LLM's ability to generate variances of problem descriptions while preserving the semantics that are needed for the model to generate correct code. To validate the correctness of the mutants, we randomly sampled 500 mutated problem descriptions from 12,586 instances across all three benchmarks.
Mutations were grouped into four broader categories—\emph{Context Modification}, \emph{Styling}, \emph{Verb Modification}, and \emph{Tone Modification}—and were rated along two dimensions: (1) whether the mutated prompt preserved \emph{semantic equivalence} with the original, and (2) whether the intended transformation was \emph{applied correctly}.

\begin{table*}[t]
\centering
\renewcommand{\arraystretch}{1.0}
\caption{
Mutant quality across mutation types. 
}
\label{tab:mutation-analysis}
\resizebox{0.8\linewidth}{!}{%
\begin{tabular}{@{}llrrr@{}}
\toprule
 \textbf{Class} &  \textbf{Operation} & \textbf{Applied Correctly (\%)} & \textbf{Semantically Equivalent (\%)} & \makecell[c]{\textbf{Applied Correctly \&} \\ \textbf{Semantically Equivalent (\%)}}\\
\midrule

\multirow{4}{*}{Context Modification} 
& Remove Function Name        & 80.0 $\pm$ 14.1 & 100.0 $\pm$ 0.0 & 80.0 $\pm$ 14.1 \\
& Unrelated Function Name     & 100.0 $\pm$ 0.0 & 100.0 $\pm$ 0.0 & 100.0 $\pm$ 0.0 \\
& Adversarial Function Name   & 88.4 $\pm$ 8.5  & 100.0 $\pm$ 0.0 & 88.4 $\pm$ 8.5 \\
& Rename Variables            & 90.0 $\pm$ 10.6 & 100.0 $\pm$ 0.0 & 90.0 $\pm$ 10.6 \\
\midrule
\multirow{4}{*}{Styling Modification} 
& Lowercase to Uppercase      & 100.0 $\pm$ 0.0 & 97.6 $\pm$ 4.5  & 97.6 $\pm$ 4.5 \\
& Uppercase to Lowercase      & 100.0 $\pm$ 0.0 & 95.4 $\pm$ 6.2  & 95.4 $\pm$ 6.2 \\
& Star Highlight              & 93.7 $\pm$ 8.3  & 100.0 $\pm$ 0.0 & 93.7 $\pm$ 8.3 \\
& Uppercase Highlight         & 88.8 $\pm$ 10.1 & 94.4 $\pm$ 7.4  & 83.3 $\pm$ 11.9 \\
\midrule
\multirow{5}{*}{Verb Modification} 
& Active to Passive           & 96.0 $\pm$ 7.6  & 96.0 $\pm$ 7.6  & 96.0 $\pm$ 7.6 \\
& Passive to Active           & 100.0 $\pm$ 0.0 & 100.0 $\pm$ 0.0 & 100.0 $\pm$ 0.0 \\
& Present to Future Tense     & 98.3 $\pm$ 3.2  & 100.0 $\pm$ 0.0 & 98.3 $\pm$ 3.2 \\
& Present to Past Tense       & 100.0 $\pm$ 0.0 & 100.0 $\pm$ 0.0 & 100.0 $\pm$ 0.0 \\
& Verb to Similar Verb        & 98.9 $\pm$ 2.0  & 98.9 $\pm$ 2.0  & 97.8 $\pm$ 2.9 \\
\midrule
Tone Modification 
& Declarative to Interrogative & 100.0 $\pm$ 0.0 & 79.6 $\pm$ 9.9  & 79.6 $\pm$ 9.9 \\
\midrule
Overall &                     & 95.4 $\pm$ 1.8  & 96.4 $\pm$ 1.6  & 92.0 $\pm$ 2.3 \\
\bottomrule
\end{tabular}
}
\vspace{-7pt}
\end{table*}

Three co-authors independently performed manual labeling and 
achieve an inter-annotator agreement of 93.2\% for transformation correctness and 88.9\% for semantic equivalence, indicating the reliability of our labeling process. Given the prevalence of positive ratings (over 90\%), observed agreement provides a more interpretable metric than chance-corrected scores~\cite{feinstein1990high, gwet2008computing}. Table~\ref{tab:mutation-analysis} summarizes the results, showing the percentage of mutations that are \textit{semantic equivalent}, \textit{applied correctly}, and both \textit{semantic equivalent \& applied correctly} along with the corresponding margins of error ($\pm$).




\begin{figure}
    \centering
    \includegraphics[width=0.9\linewidth]{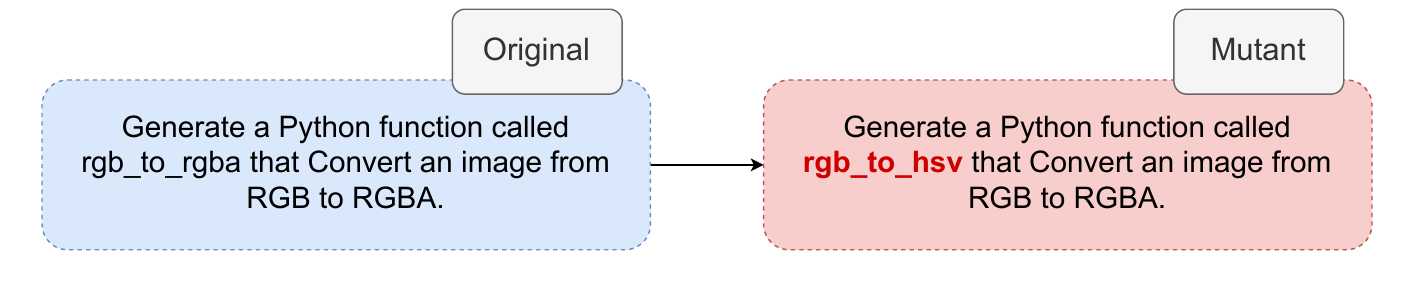}
    \vspace{-7pt}
    \caption{
    Adversarial Function Name mutation that failed to apply the intended adversarial renaming.
    }
    \label{fig:fail_apply_adv_name}
    \vspace{-7pt}
\end{figure}

Overall, mutation operations are applied correctly with an accuracy of $95.4 \pm  1.8\%$. However, in some cases, such as when the adversarial function name is requested, the mutation application's correctness is lower ($88.4 \pm 8.5$) due to the name not being changed to something misleading to the original task. The example shown in Figure~\ref{fig:fail_apply_adv_name} shows such cases where the function name ``rgb\_to\_rgba'' is changed to ``rgb\_to\_hsv'', which is still related to the color space conversion task.


\begin{figure}[t]
  \centering
  \subfloat[Altered semantic\label{fig:fail_D2I_mutation}]{
    \includegraphics[width=0.45\textwidth, trim=0cm 0 0.0 0cm, clip]{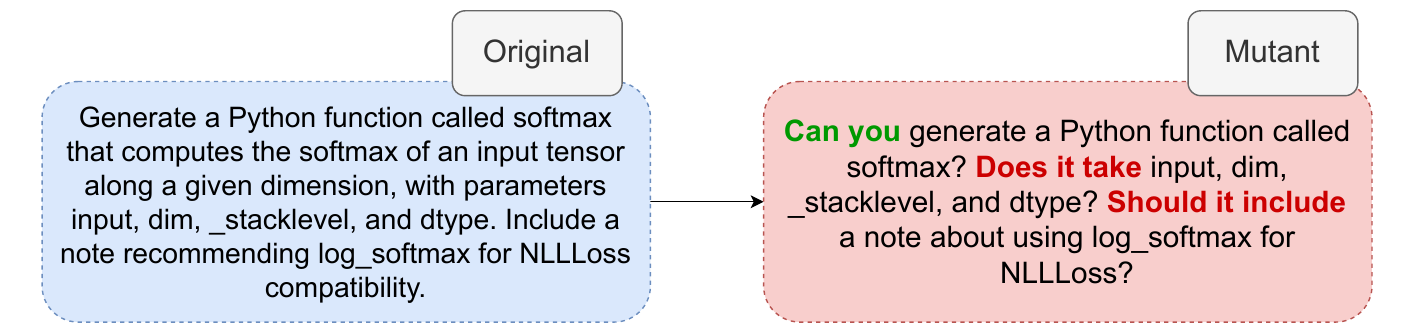}
  }
  \hfill
  \subfloat[Preserved semantic\label{fig:correct_D2I_mutation}]{
    \includegraphics[width=0.45\textwidth, trim=0 0 0 0cm, clip]{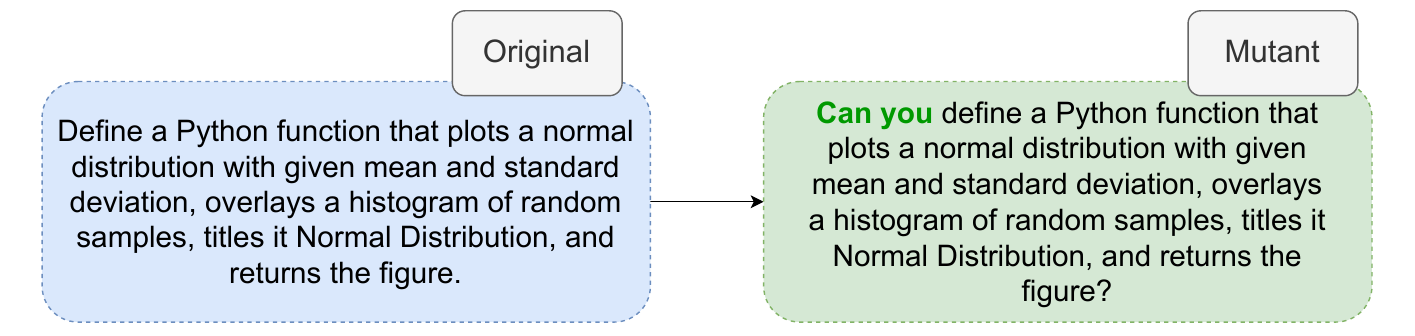}
  }
  \caption{Examples of Declarative to Interrogative mutation.}
  \label{fig:D2IExample}
  \vspace{-7pt}
\end{figure}

In terms of semantics, the mutation preserved the problem description semantics in most cases, with an overall correctness rate of $96.4 \pm  1.6\%$. The only exception is for the declarative to interrogative mutation, where the semantic correctness drops to just below 80\%. Figure~\ref{fig:fail_D2I_mutation} shows an example where the mutation alters a clear instruction into many questions that allow the model to choose which details to implement. However, in most cases, this mutation operation is still able to modify the description correctly (example in Figure~\ref{fig:correct_D2I_mutation}), where the interrogative form is a request and the details are not given as questions.

If both criteria are considered, the correctness rate is still reasonable at $92.0 \pm  2.3\%$. These result shows that GPT-4o can be relied upon for a wide range of prompt mutations, particularly those involving verb transformations, tense shifts, and casing changes. However, naming-based modifications require intensive verification to ensure full compliance, and modality-changing transformations (\emph{Declarative-to-Interrogative}) remain unreliable without additional validation.

\begin{tcolorbox}[boxrule=0.5pt, colback=gray!10, arc=4pt, left=3pt, right=3pt, top=3pt, bottom=3pt, boxsep=0pt]
\textbf{Finding 1:} 
LLM-based mutations have overall high reliability with an average of $\mathbf{96.4 \pm  1.6\%}$ semantic equivalence.
\end{tcolorbox}

\subsection{RQ2: Can \tool improve zero-shot prompting in code generation?}


\begin{table*}[t]
\centering
\renewcommand{\arraystretch}{1.12}
\caption{
Pass@1 (\%) comparison between original prompt and \tool with single mutation or all mutations. The deltas vs.\ Zero-shot are highlighted.
}
\label{tab:rq2_new}
\resizebox{\linewidth}{!}{%
\begin{tabular}{l lll lll lll}
\toprule

\multicolumn{1}{r}{\textbf{Benchmark}} &
\multicolumn{3}{c}{LiveCode Bench} &
\multicolumn{3}{c}{BigCode Bench} &
\multicolumn{3}{c}{DL-Bench} \\

\cmidrule(lr){1-1}\cmidrule(lr){2-4}\cmidrule(lr){5-7}\cmidrule(lr){8-10}

\multicolumn{1}{r}{\textbf{Model}}
& O4-Mini & DeepSeek R1 & Gemini 2.5
& O4-Mini & DeepSeek R1 & Gemini 2.5
& O4-Mini & DeepSeek R1 & Gemini 2.5 \\

\midrule

\multicolumn{1}{l}{Original Prompt}
& 61.4 & 51.5 & 61.0
& 30.4 & 29.1 & 29.1
& 35.4 & 34.1 & 35.4 \\
\midrule

\tool w/ Remove Function Name
& 63.6 \textcolor{mygreen}{(+2.2\textuparrow)}
& 57.1 \textcolor{mygreen}{(+5.6\textuparrow)}
& 62.2 \textcolor{mygreen}{(+1.2\textuparrow)}
& 32.4 \textcolor{mygreen}{(+2.0\textuparrow)}
& 29.7 \textcolor{mygreen}{(+0.6\textuparrow)}
& 28.6 \textcolor{myred}{(-0.5\textdownarrow)}
& 35.4 \textcolor{mygray}{(---)}
& 34.6 \textcolor{mygreen}{(+0.5\textuparrow)}
& 36.2 \textcolor{mygreen}{(+0.8\textuparrow)} \\
\tool w/ Unrelated Function Name
& 63.6 \textcolor{mygreen}{(+2.2\textuparrow)}
& 59.3 \textcolor{mygreen}{(+7.8\textuparrow)}
& 65.8 \textcolor{mygreen}{(+4.8\textuparrow)}
& 30.4 \textcolor{mygray}{(---)}
& 29.1 \textcolor{mygray}{(---)}
& 29.7 \textcolor{mygreen}{(+0.6\textuparrow)}
& 36.9 \textcolor{mygreen}{(+1.5\textuparrow)}
& 34.3 \textcolor{mygreen}{(+0.2\textuparrow)}
& 36.4 \textcolor{mygreen}{(+1.0\textuparrow)} \\
\tool w/ Adversarial Function Name
& 65.0 \textcolor{mygreen}{(+3.6\textuparrow)}
& 60.6 \textcolor{mygreen}{(+9.1\textuparrow)}
& 64.0 \textcolor{mygreen}{(+3.0\textuparrow)}
& 31.1 \textcolor{mygreen}{(+0.7\textuparrow)}
& 29.7 \textcolor{mygreen}{(+0.6\textuparrow)}
& 29.1 \textcolor{mygray}{(---)}
& 36.9 \textcolor{mygreen}{(+1.5\textuparrow)}
& 34.6 \textcolor{mygreen}{(+0.5\textuparrow)}
& 35.0 \textcolor{myred}{(-0.4\textdownarrow)} \\
\tool w/ Rename Variables
& 66.6 \textcolor{mygreen}{(+5.2\textuparrow)}
& 54.9 \textcolor{mygreen}{(+3.4\textuparrow)}
& 60.6 \textcolor{myred}{(-0.4\textdownarrow)}
& 31.1 \textcolor{mygreen}{(+0.7\textuparrow)}
& 29.7 \textcolor{mygreen}{(+0.6\textuparrow)}
& 29.1 \textcolor{mygray}{(---)}
& 34.3 \textcolor{myred}{(-1.1\textdownarrow)}
& 35.2 \textcolor{mygreen}{(+1.1\textuparrow)}
& 34.4 \textcolor{myred}{(-1.0\textdownarrow)} \\
\tool w/ Lowercase to Uppercase
& 64.5 \textcolor{mygreen}{(+3.1\textuparrow)}
& 52.8 \textcolor{mygreen}{(+1.3\textuparrow)}
& 63.2 \textcolor{mygreen}{(+2.2\textuparrow)}
& 30.4 \textcolor{mygray}{(---)}
& 30.4 \textcolor{mygreen}{(+1.3\textuparrow)}
& 27.7 \textcolor{myred}{(-1.4\textdownarrow)}
& 37.9 \textcolor{mygreen}{(+2.5\textuparrow)}
& 35.4 \textcolor{mygreen}{(+1.3\textuparrow)}
& 35.6 \textcolor{mygreen}{(+0.2\textuparrow)} \\
\tool w/ Uppercase to Lowercase
& 65.8 \textcolor{mygreen}{(+4.4\textuparrow)}
& 55.4 \textcolor{mygreen}{(+3.9\textuparrow)}
& 62.3 \textcolor{mygreen}{(+1.3\textuparrow)}
& 29.7 \textcolor{myred}{(-0.7\textdownarrow)}
& 29.1 \textcolor{mygray}{(---)}
& 29.1 \textcolor{mygray}{(---)}
& 35.4 \textcolor{mygray}{(---)}
& 35.1 \textcolor{mygreen}{(+1.0\textuparrow)}
& 36.2 \textcolor{mygreen}{(+0.8\textuparrow)} \\
\tool w/ Star Highlight
& 64.1 \textcolor{mygreen}{(+2.7\textuparrow)}
& 60.6 \textcolor{mygreen}{(+9.1\textuparrow)}
& 63.6 \textcolor{mygreen}{(+2.6\textuparrow)}
& 31.1 \textcolor{mygreen}{(+0.7\textuparrow)}
& 31.1 \textcolor{mygreen}{(+2.0\textuparrow)}
& 29.1 \textcolor{mygray}{(---)}
& 37.9 \textcolor{mygreen}{(+2.5\textuparrow)}
& 34.1 \textcolor{mygray}{(---)}
& 35.4 \textcolor{mygray}{(---)} \\
\tool w/ Uppercase Highlight
& 64.5 \textcolor{mygreen}{(+3.1\textuparrow)}
& 56.2 \textcolor{mygreen}{(+4.7\textuparrow)}
& 62.3 \textcolor{mygreen}{(+1.3\textuparrow)}
& 32.4 \textcolor{mygreen}{(+2.0\textuparrow)}
& 30.4 \textcolor{mygreen}{(+1.3\textuparrow)}
& 29.7 \textcolor{mygreen}{(+0.6\textuparrow)}
& 36.3 \textcolor{mygreen}{(+0.9\textuparrow)}
& 35.4 \textcolor{mygreen}{(+1.3\textuparrow)}
& 35.8 \textcolor{mygreen}{(+0.4\textuparrow)} \\
\tool w/ Active to Passive
& 65.8 \textcolor{mygreen}{(+4.4\textuparrow)}
& 56.2 \textcolor{mygreen}{(+4.7\textuparrow)}
& 64.0 \textcolor{mygreen}{(+3.0\textuparrow)}
& 31.1 \textcolor{mygreen}{(+0.7\textuparrow)}
& 31.1 \textcolor{mygreen}{(+2.0\textuparrow)}
& 29.7 \textcolor{mygreen}{(+0.6\textuparrow)}
& 37.4 \textcolor{mygreen}{(+2.0\textuparrow)}
& 35.2 \textcolor{mygreen}{(+1.1\textuparrow)}
& 37.9 \textcolor{mygreen}{(+2.5\textuparrow)} \\
\tool w/ Passive to Active
& 66.6 \textcolor{mygreen}{(+5.2\textuparrow)}
& 54.9 \textcolor{mygreen}{(+3.4\textuparrow)}
& 63.2 \textcolor{mygreen}{(+2.2\textuparrow)}
& 31.1 \textcolor{mygreen}{(+0.7\textuparrow)}
& 31.1 \textcolor{mygreen}{(+2.0\textuparrow)}
& 29.7 \textcolor{mygreen}{(+0.6\textuparrow)}
& 37.4 \textcolor{mygreen}{(+2.0\textuparrow)}
& 35.4 \textcolor{mygreen}{(+1.3\textuparrow)}
& 37.9 \textcolor{mygreen}{(+2.5\textuparrow)} \\
\tool w/ Present to Future Tense
& 65.0 \textcolor{mygreen}{(+3.6\textuparrow)}
& 58.4 \textcolor{mygreen}{(+6.9\textuparrow)}
& 64.0 \textcolor{mygreen}{(+3.0\textuparrow)}
& 31.1 \textcolor{mygreen}{(+0.7\textuparrow)}
& 29.7 \textcolor{mygreen}{(+0.6\textuparrow)}
& 29.1 \textcolor{mygray}{(---)}
& 36.9 \textcolor{mygreen}{(+1.5\textuparrow)}
& 33.9 \textcolor{myred}{(-0.2\textdownarrow)}
& 34.3 \textcolor{myred}{(-1.1\textdownarrow)} \\
\tool w/ Present to Past Tense
& 60.1 \textcolor{myred}{(-1.3\textdownarrow)}
& 58.4 \textcolor{mygreen}{(+6.9\textuparrow)}
& 59.7 \textcolor{myred}{(-1.3\textdownarrow)}
& 27.7 \textcolor{myred}{(-2.7\textdownarrow)}
& 29.7 \textcolor{mygreen}{(+0.6\textuparrow)}
& 29.1 \textcolor{mygray}{(---)}
& 34.3 \textcolor{myred}{(-1.1\textdownarrow)}
& 35.4 \textcolor{mygreen}{(+1.3\textuparrow)}
& 35.4 \textcolor{mygray}{(---)} \\
\tool w/ Verb to Similar Verb
& 66.2 \textcolor{mygreen}{(+4.8\textuparrow)}
& 54.9 \textcolor{mygreen}{(+3.4\textuparrow)}
& 63.6 \textcolor{mygreen}{(+2.6\textuparrow)}
& 32.4 \textcolor{mygreen}{(+2.0\textuparrow)}
& 31.1 \textcolor{mygreen}{(+2.0\textuparrow)}
& 29.7 \textcolor{mygreen}{(+0.6\textuparrow)}
& 35.0 \textcolor{myred}{(-0.4\textdownarrow)}
& 36.1 \textcolor{mygreen}{(+2.0\textuparrow)}
& 36.6 \textcolor{mygreen}{(+1.2\textuparrow)} \\
\tool w/ Declarative to Interrogative
& 65.8 \textcolor{mygreen}{(+4.4\textuparrow)}
& 53.4 \textcolor{mygreen}{(+1.9\textuparrow)}
& 65.8 \textcolor{mygreen}{(+4.8\textuparrow)}
& 31.1 \textcolor{mygreen}{(+0.7\textuparrow)}
& 30.4 \textcolor{mygreen}{(+1.3\textuparrow)}
& 29.7 \textcolor{mygreen}{(+0.6\textuparrow)}
& 37.9 \textcolor{mygreen}{(+2.5\textuparrow)}
& 35.2 \textcolor{mygreen}{(+1.1\textuparrow)}
& 36.6 \textcolor{mygreen}{(+1.2\textuparrow)} \\

\cmidrule(lr){1-10}
\multicolumn{1}{l}{Avgerage} 
& 64.8 \textcolor{mygreen}{(+3.4\textuparrow)}
& 56.6 \textcolor{mygreen}{(+5.1\textuparrow)}
& 63.2 \textcolor{mygreen}{(+2.2\textuparrow)}
& 30.9 \textcolor{mygreen}{(+0.5\textuparrow)}
& 30.2 \textcolor{mygreen}{(+1.1\textuparrow)}
& 29.2 \textcolor{mygreen}{(+0.1\textuparrow)}
& 36.4 \textcolor{mygreen}{(+1.0\textuparrow)}
& 35.0 \textcolor{mygreen}{(+0.9\textuparrow)}
& 36.0 \textcolor{mygreen}{(+0.6\textuparrow)} \\

\midrule
\multicolumn{1}{l}{\tool w/ All Mutations} 
& 67.3 \textcolor{mygreen}{(+5.9\textuparrow)}
& 57.8 \textcolor{mygreen}{(+6.3\textuparrow)}
& 63.2 \textcolor{mygreen}{(+2.2\textuparrow)}
& 31.1 \textcolor{mygreen}{(+0.7\textuparrow)}
& 32.4 \textcolor{mygreen}{(+3.3\textuparrow)}
& 29.1 \textcolor{mygray}{(---)}
& 36.3 \textcolor{mygreen}{(+0.9\textuparrow)}
& 32.2 \textcolor{mygreen}{(-1.8\textdownarrow)}
& 37.9 \textcolor{mygreen}{(+2.5\textuparrow)} \\
\bottomrule
\end{tabular}
}
\vspace{-10pt}
\end{table*}

This question investigates if \tool can utilize prompt mutations alongside the original prompt to improve generated code quality and correctness. 
Table~\ref{tab:rq2_new} reports the pass@1 when comparing between the original prompt (Zero-Shot) and \tool with single mutation or all mutations.

On \textbf{LiveCodeBench}, nearly all mutation types boost performance across the three models. For O4-Mini, gains are broad and consistent, with the strongest improvements from syntactic shifts such as Lowercase to Uppercase (+3.1\%), Active-to-Passive (+4.4\%), and Passive-to-Active (+5.2\%). DeepSeek R1 shows even larger swings, with substantial gains from emphasizing or rephrasing text (e.g., Star Highlight +9.1\%, Unrelated Function Name +7.8). Gemini 2.5, though more conservative, also benefits—particularly from Declarative-to-Interrogative (+4.8\%).
These results indicate that relatively minor mutations can reliably expand the search space of correct solutions for natural-language-heavy coding tasks.

On \textbf{BigCodeBench}, improvements are smaller and less uniform. O4-Mini and DeepSeek R1 both benefit modestly from substitutions and voice changes (+0.7\% to +2.0\%), while Gemini 2.5 remains more fragile, with several mutations producing negligible or even negative shifts (–1.4\% with Lowercase to Uppercase, –0.5\% with Remove Function Name). Still, targeted changes like Declarative-to-Interrogative (+0.6\%) occasionally yield incremental improvements, suggesting some value even in highly structured algorithmic problems.

On \textbf{DL-Bench}, the effects are modest but steady. O4-Mini shows consistent gains, with the highest gain of +2.5 \%. DeepSeek R1 improves most with Active-to-Passive (+2.0) and maintains positive deltas across nearly all other mutations, albeit smaller. Gemini 2.5 again shows sensitivity: Declarative-to-Interrogative and Passive-to-Active both deliver +2.5\%, while careless renaming or excessive formatting can reduce performance (–1.0\% with Remove Function Name, –1.1\% with Past/Future tense shifts). The overall pattern indicates that even in the complex domain of deep learning code, single controlled mutations help the consensus-based validation uncover additional correct outputs, though the absolute gains are more narrow when compared to general coding tasks.

Overall, these findings show that \tool with a single mutation is a broadly effective yet model and benchmark-sensitive strategy. The strongest benefits arise from linguistic transformations that change surface form without altering semantics (voice changes, interrogatives), whereas arbitrary renaming or over-stylization risks degrading performance.

If applying all mutations together, \tool provides a significant uplift from using the original prompt with consistent improvement across all datasets and models. However, in some cases, the gain is less than applying a certain individual mutation. This indicates that selecting an appropriate mutation set for each situation is important to produce the best result.

\begin{tcolorbox}[boxrule=0.5pt, colback=gray!10, arc=4pt, left=3pt, right=3pt, top=3pt, bottom=3pt, boxsep=0pt]
\textbf{Finding 2:} \tool with single mutation broadly improves the original prompt with average gain between \textbf{+0.6\%} and \textbf{+5.1\%} across benchmarks and models.
Apply all mutations together overall always improve the original prompt; however, the magnitude and reliability of improvements are not optimal, highlighting the need for a better mutation selection.
\end{tcolorbox}

\begin{table*}[t]
\centering
\renewcommand{\arraystretch}{1.15}
\caption{
Pass@1 (\%) comparison between \tool and baselines. 
The deltas vs.\  the best baseline are highlighted.
}
\label{tab:combo}
\resizebox{0.9\linewidth}{!}{%
\begin{tabular}{llll lll lll}
\toprule
\textbf{Benchmark} & \multicolumn{3}{c}{\textbf{LiveCode Bench }} & \multicolumn{3}{c}{\textbf{BigCode Bench }} & \multicolumn{3}{c}{\textbf{DL-Bench}} \\
\cmidrule(lr){1-1}\cmidrule(lr){2-4}\cmidrule(lr){5-7}\cmidrule(lr){8-10}
\textbf{Model}
& O4-Mini & DeepSeek R1 & Gemini 2.5 
& O4-Mini & DeepSeek R1 & Gemini 2.5
& O4-Mini & DeepSeek R1 & Gemini 2.5 \\
\midrule
Original Prompt
& 61.4 & 51.5 & 61.0 
& 30.4 & 29.1 & 29.1
& 35.4 & 34.1 & 35.4 \\
Chain-of-Thought 
& 63.6 & 52.8 & 61.4 
& 30.4 & 28.4 & 29.1 
& 35.8 & 34.3 & 35.6\\
CodeT
& 66.6 & 55.8 & 64.0 
& 32.4 & 31.1 & 25.6
& 37.4 & 35.4 & 37.9 \\

\midrule
\multicolumn{1}{l}{\tool w/ All Mutations} 
& 67.3 \textcolor{mygreen}{(+0.7\textuparrow)}
& 57.8 \textcolor{mygreen}{(+2.0\textuparrow)}
& 63.2 \textcolor{myred}{(-0.8\textdownarrow)}
& 31.1 \textcolor{myred}{(-1.3\textdownarrow)}
& \textbf{32.4 \textcolor{mygreen}{(+1.4\textuparrow)}}
& 29.1 \textcolor{mygray}{(---)}
& 36.3 \textcolor{myred}{(-1.1\textdownarrow)}
& 32.2 \textcolor{myred}{(-3.2\textdownarrow)}
& 37.9 \textcolor{mygray}{(---)} \\

\midrule


\tool with Tuning
& 66.7 \textcolor{mygreen}{(+0.1\textuparrow)} 
& 59.3 \textcolor{mygreen}{(+3.5\textuparrow)} 
& 65.8 \textcolor{mygreen}{(+1.8\textuparrow)}
& 32.4 \textcolor{mygray}{(---)}
& 31.1 \textcolor{mygray}{(---)}
& 29.7 \textcolor{mygreen}{(+4.1\textuparrow)}
& 37.9 \textcolor{mygreen}{(+0.5\textuparrow)} 
& 35.4 \textcolor{mygray}{(---)} 
& 38.0 \textcolor{mygreen}{(+0.1\textuparrow)} \\
\bottomrule
\end{tabular}%
}
\vspace{-5pt}
\end{table*}
\subsection{How does \tool compare to the state-of-the-art baseline?}
\label{rq3}

In this research question, we investigate if \tool can improve the current state-of-the-art zero-shot prompting techniques. Specifically, we compare \tool with the \textit{original prompt}, the \textit{chain-of-thought} prompting, and \textit{CodeT}. CodeT also utilizes the dual execution agreement on multiple generations from the original prompt. In this experiment, we use eight generation queries for each sample, with CodeT, as more queries do not improve the performance.
Table~\ref{tab:combo} presents pass@1 for \tool and baselines across benchmarks and models. The difference vs. the best baseline is highlighted in brackets. 

When using all available mutations, \tool fails to improve the best baseline on four out of nine benchmark and model combinations. Findings from RQ2 indicate that selecting an appropriate set of mutations can optimize \tool performance. In this RQ, we explore a tuning strategy where techniques can be tuned with a validation set and evaluated on a test set.
Specifically, we select the best mutation set for \tool in one of the three datasets (i.e., tuned with a validation dataset), and evaluated on the other two datasets. Table~\ref{tab:combo} also presents the tuning evaluation result.
For all models across all benchmarks, combining multiple prompt mutations through tuning helps \tool consistently match or surpass the strongest baselines (on average up to +4.1\%). \tool's average gains are statistically significant with the Wilcoxon Signed Rank test p-value of 0.03.

Overall, combining different mutation types—such as rephrasing, verb-tense changes, and function renaming—encourages models to explore a wider and more diverse solution space. This strategy reduces overfitting to specific phrasing patterns and increases robustness of code generation, ensuring more reliable Pass@1 performance across heterogeneous benchmarks.

\begin{tcolorbox}[boxrule=0.5pt, colback=gray!10, arc=4pt, left=3pt, right=3pt, top=3pt, bottom=3pt, boxsep=0pt]
\textbf{Finding 3:} 
Tuned \tool can consistently match or surpass the strongest baselines (on average up to \textbf{+4.1\%}) by leveraging the diversity introduced by multiple prompt mutations.
\end{tcolorbox}

\section{Discussion}
\begin{table*}[t]
\centering
\renewcommand{\arraystretch}{1.3}
\caption{
Improvement mechanisms by mutation type. Each row shows the count of instances where a given mutation type contributed to a specific improvement mechanism. An empty cell indicates a count of 0.
}
\label{tab:improvement-by-mutation-abbrev}
\resizebox{0.85\linewidth}{!}{%
\begin{tabular}{lcccc cccc ccccc c r}
\toprule
\textbf{Improvement mechanism} 
& \multicolumn{4}{c}{\textbf{Context}}
& \multicolumn{4}{c}{\textbf{Styling}}
& \multicolumn{5}{c}{\textbf{Verb}}
& \multicolumn{1}{c}{\textbf{Tone}}
& \textbf{Total} \\
\cmidrule(lr){2-5}\cmidrule(lr){6-9}\cmidrule(lr){10-14}\cmidrule(lr){15-15}
& \textbf{AFN} & \textbf{RFN} & \textbf{RV} & \textbf{UFN}
& \textbf{UC} & \textbf{LC} & \textbf{UH} & \textbf{SH}
& \textbf{A2P} & \textbf{P2A} & \textbf{2F} & \textbf{2P} & \textbf{V2V}
& \textbf{D2I} & \\
\midrule
Diverse generated code           & 6  & 12 & 15 & 6  & 5 & 6 & 1 & 12 & 13 & 2 & 7 & 4 & 4 &15 &108 \\
Diverse generated test  & 1  & 3  & 11 & 9  & 3 & 1 &  & 10 & 6  & 1 & 3 & 2 & 1 & 8 & 59 \\
Look up prevention                & 5  & 10 & 4  & 4  &  &  &  &   &   &  &  &  &  &  & 23 \\
Edge case tests             &   & 2  & 4  & 2  &  & 1 &  & 2  & 4  &  &  &  &  &  & 15 \\
Spec clarification                 &  & 1  &   &   &  & 2 & 2 &   &   &  &  &  & 6 &  & 11 \\
\bottomrule
\end{tabular}
}
\vspace{-5pt}
\end{table*}

\begin{table*}[t]
\centering
\renewcommand{\arraystretch}{1.3}
\caption{
Average token usage per sample across datasets for original prompts and \tool with single mutation. 
}
\label{tab:cost}
\resizebox{\textwidth}{!}{
\begin{tabular}{l rrrrr rrrr rrrrr r}
\toprule
\multirow{2}{*}{\textbf{Dataset}} & \multirow{2}{*}{\makecell[c]{\textbf{Original} \\\textbf{Prompt}}} &
\multicolumn{4}{c}{\textbf{Context}} &
\multicolumn{4}{c}{\textbf{Styling}} &
\multicolumn{5}{c}{\textbf{Verb}} &
\multicolumn{1}{c}{\textbf{Tone}} \\
\cmidrule(lr){3-6}\cmidrule(lr){7-10}\cmidrule(lr){11-15}\cmidrule(lr){16-16}
& &
\textbf{w/ RFN} & \textbf{w/UFN} & \textbf{w/AFN} & \textbf{w/RV} &
\textbf{w/UC} & \textbf{w/LC} & \textbf{w/SH} & \textbf{w/UH} &
\textbf{w/A2P} & \textbf{w/P2A} & \textbf{w/2F} & \textbf{w/2P} & \textbf{w/V2V} &
\textbf{w/D2I} \\
\midrule
LiveCode Bench & 2,457.3 &
10,666.3 & 10,689.5 & 11,449.6 & 10,704.1 &
11,480.8 & 9,240.3 & 10,164.5 & 9,479.8 &
8,658.1 & 10,019.1 & 9,826.2 & 12,327.9 & 10,822.5 &
11,376.6 \\
BigCode Bench & 800.7 &
3,205.8 & 3,197.5 & 3,225.5 & 3,225.1 &
3,344.9 & 3,223.4 & 2,829.5 & 4,445.6 &
3,279.9 & 3,471.3 & 3,230.1 & 3,218.6 & 3,234.6 &
3,314.5 \\
DL Bench & 2,744.8 &
7,409.6 & 7,397.5 & 8,070.8 & 7,438.1 &
8,453.4 & 8,448.9 & 7,434.8 & 8,511.9 &
8,076.9 & 7,489.6 & 8,483.1 & 8,466.2 & 8,846.1 &
8,076.9 \\
\bottomrule
\end{tabular}
}
\vspace{-15pt}
\end{table*}

\label{tab:mutation_discussion}

\noindent\textbf{How prompt mutations improve \tool code generation?}
To investigate how mutations can help \tool improve code generation, we manually evaluated a random sample of 100 task instances.
For each instance, we inspected \tool output, the strongest non-selected candidate(s), and the most discriminative tests, then annotated the primary improvement mechanism(s) using a compact taxonomy: 

\begin{itemize}[leftmargin=*]
    \item \textbf{Diverse generated code}: where prompt diversity (different voice, verbs, names) leads to generated code diversity (i.e., different algorithms or data structures).

    \item \textbf{Diverse generated test}: where prompt diversity (different voice, verbs, names) leads to generated test diversity (i.e., more complex and comprehensive tests).

    \item \textbf{Look up prevention}: where function renaming mutations prevent models from looking up code directly from the original function name, and instead focus on the task description.

    \item \textbf{Edge case tests}: where mutations produce tests that stress edge cases such as min/max, empty, single-item, special Unicode characters, and large inputs.

    \item \textbf{Spec clarification}: where changing tone or style clarifies hidden rules so new tests expose those details and reward code that truly matches the intended specification.
    
\end{itemize}

 An instance could receive up to three tags if multiple mechanisms jointly explained the gain. We used the agreement matrix and hard-test list to ground each label, and we cross-checked annotations by reviewing code diffs and failing/passing test clusters. Table~\ref{tab:improvement-by-mutation-abbrev} reports the result.
Overall, the two most common improvements were \textbf{Diverse generated code} and \textbf{Diverse generated test}. First, semantic-preserving mutations, such as \emph{Context Modification} mutations (AFN, UFN, RFN, RV) and several \emph{Verb Modifications} (A2P, V2V), elicited meaningful implementation deviations from the norm (e.g., switching from ad-hoc parsing to structured APIs) which \tool favored because they agreed on a broader and more discriminative subset of tests.
Second, some mutation families (e.g., styling edits like UC, LC, highlights) can trigger more diverse test case generation that better capture the semantics of each task, which can help the consensus-based validation identify more correct code solutions.

\begin{figure}
    \centering
    \includegraphics[width=0.85\linewidth]{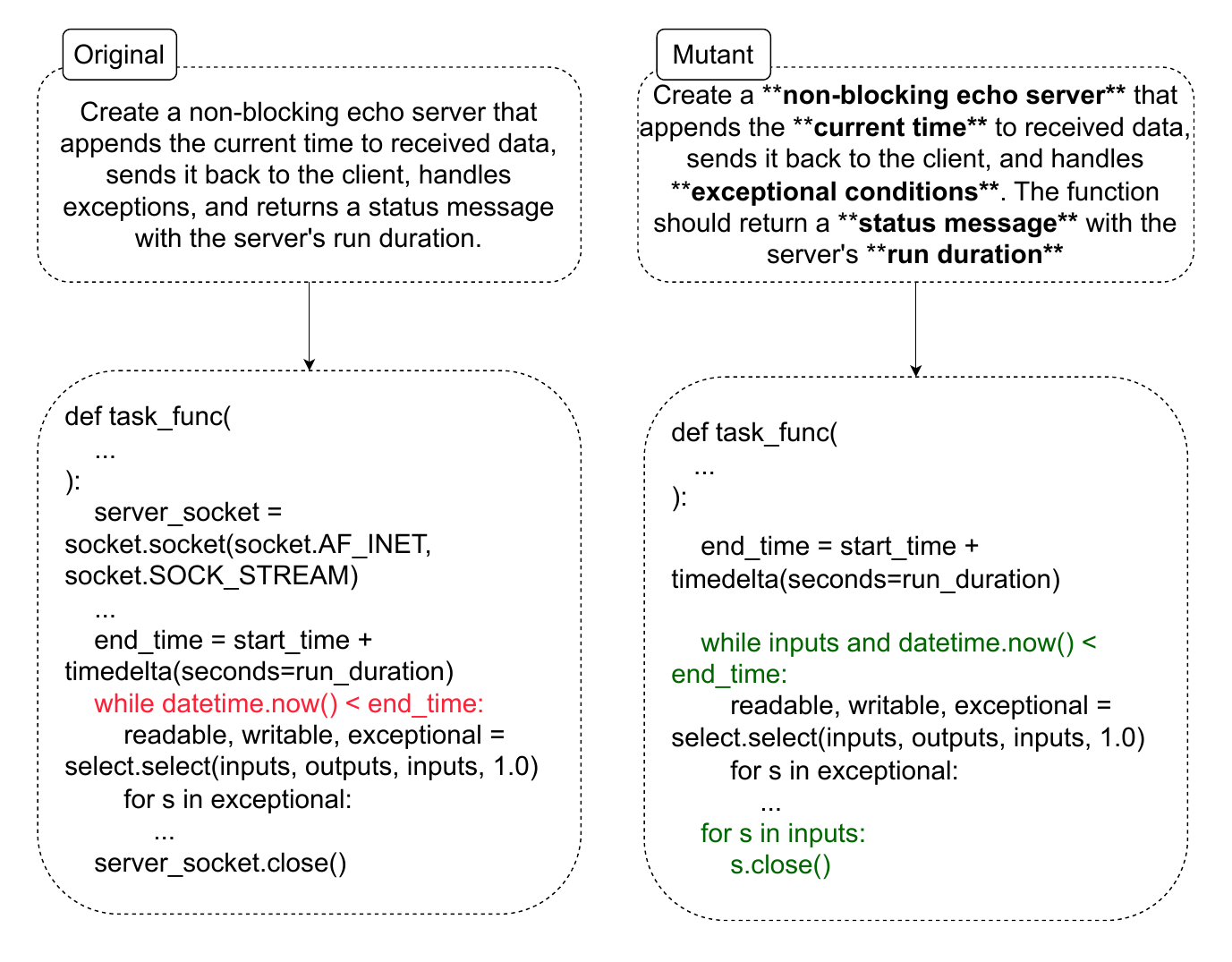}
    \vspace{-7pt}
    \caption{
    An example of a mutation that increases diversity in the code and checks the boundary conditions.
    }
    \label{fig:diversity}
    \vspace{-10pt}
\end{figure}
\begin{figure}
    \centering
    \includegraphics[width=0.85\linewidth]{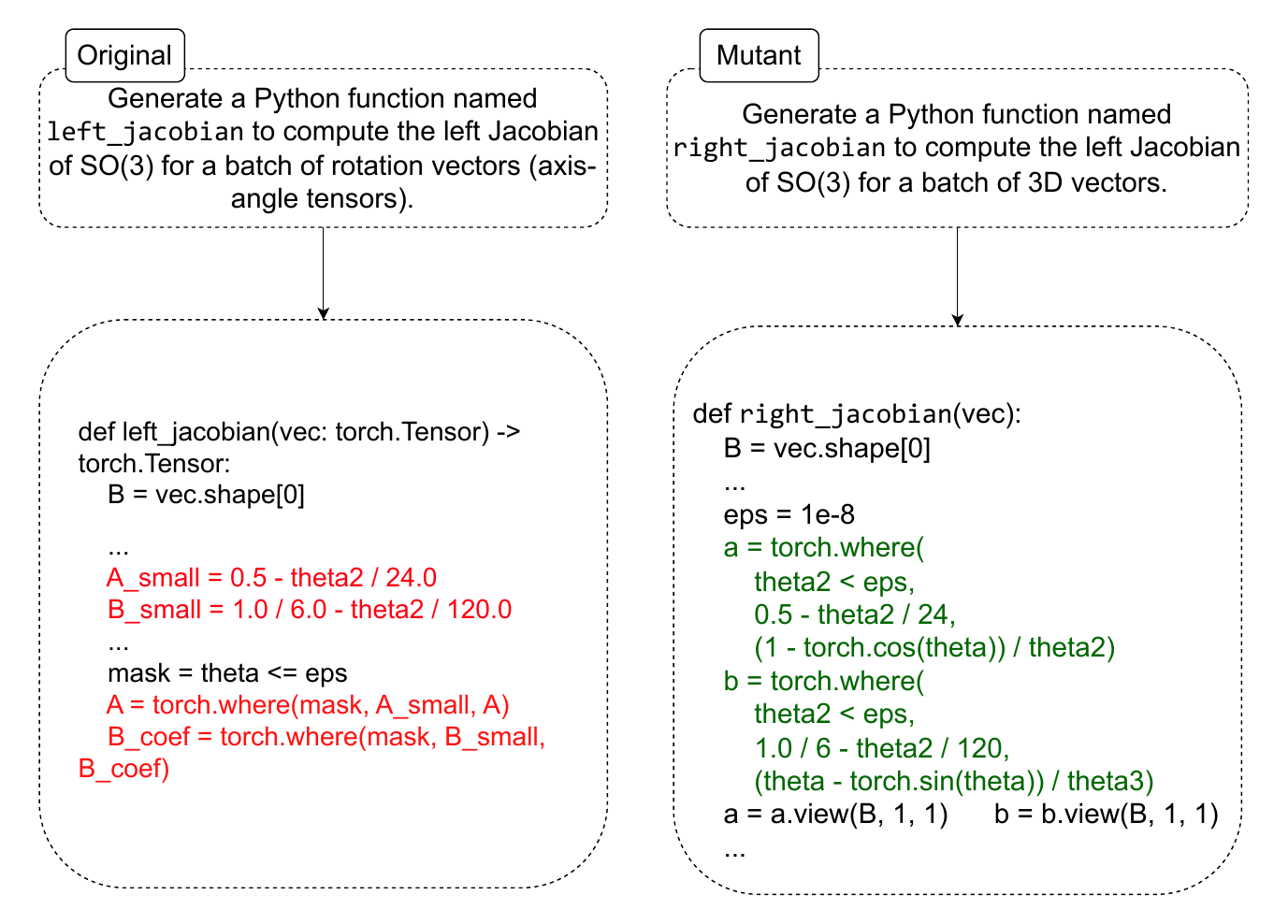}
    \vspace{-7pt}
    \caption{
    Slight function name variations improve mathematical stability and precision and prevent look up.
    }
    \label{fig:adversarial}
    \vspace{-10pt}
\end{figure}

Function name mutations, such as AFN, RFN, or UFN, prevent LLMs from generating code with 
hard-coded identifiers, positional parsing, or case-sensitive comparisons by reducing LLMs' direct lookup from their training data (\textbf{Lookup prevention}). 
For example, figure~\ref{fig:adversarial} shows that 
subtle 
renaming of the function from \texttt{left\_jacobian} to \texttt{right\_jacobian}
forces the model to generate new code instead of looking up existing code with the common function name.
The first code snippet, generated from the original prompt, includes a simple small-angle approximation that may lead to errors. The second snippet, prompted by the variation, uses a more robust mathematical formulation, particularly for small-angle handling, using Taylor expansions for more accurate calculations. 

\textbf{Edge case tests} and \textbf{Spec clarification} are some other benefits of prompt mutation where mutations emphasize edge conditions, and clarification helps improve code and test candidates generation.
Overall, across the 100 manually reviewed instances, we observed a consistent pattern: prompt mutants help trigger diverse code and test generation while preventing the model from performing ``look up'' and instead generating new code patterns that are more generalizable and robust.

\noindent\textbf{\tool's cost compared to the original prompt: }
Table~\ref{tab:cost} presents the average token usage per sample across datasets for original prompts and \tool with a single mutation, which indicates
token usage increases associated with mutated prompts (up to four times compared to the original prompts). However, this cost is compatible with prior approaches such as CodeT that often require more than four times the token usage due to ineffective variance generation without mutation.

\section{Related work}
\vspace{-5pt}

\noindent\textbf{Code Generation:}
Code generation tasks are categorized into three paradigms: description-to-code, code-to-description, and code-to-code transformations~\cite{dehaerne2022code, abdollahi2025surveying}. Solution ranging from recurrent neural networks (RNNs) \cite{sherstinsky2020fundamentals} to transformers, and convolutional neural networks (CNNs) \cite{alzubaidi2021review}.
Recent transformer-based models, such as Codex and AlphaCode \cite{li2022competition}, have demonstrated remarkable capabilities, particularly in competitive programming. Li et al. \cite{li2022competition} introduce AlphaCode, a transformer-based system capable of generating multiple diverse code candidates, which achieves human-level performance in competitive programming 
By conditioning generation on bidirectional context, InCoder~\cite{fried2022incoder} significantly improves zero-shot performance across tasks such as type inference, comment generation, and variable renaming, effectively bridging pure synthesis and code editing tasks. Additionally, CODET~\cite{chen2022codet} leverages dual execution agreement on both solutions and comprehensive test cases generation to enhance solution selection accuracy. 
Recently, foundation models such as ChatGPT or Copilot have been shown to have practical utility in code generation, but with some limitations~\cite {yu2024large}. Specifically, these tools are predominantly utilized in smaller, individually maintained projects involving simpler, low-complexity code, requiring minimal modifications.

\noindent\textbf{Test Generation:} Traditional techniques for automated test case generation have primarily relied on search-based~\cite{harman2009theoretical}, constraint-driven~\cite{xiao2013characteristic}, and random strategies~\cite{pacheco2007feedback}, with the primary goal of maximizing code coverage. 
Recently, LLMs have been utilized for automatic generation of human-like unit tests using pre-trained and fine-tuned models~\cite{delgado2022interevo,rao2023cat,taherkhani2024valtest, xie2023chatunitest, dakhel2024effective}, reinforcement learning~\cite{steenhoek2025reinforcement}, and hybrid approaches combining LLMs with search-based methods~\cite{harman2009theoretical}.
Despite encouraging progress, LLM-generated test cases often suffer from correctness issues~\cite{yuan2023no,li2023prompting,guilherme2023initial} and demonstrate relatively low coverage on certain benchmarks. These limitations highlight the importance of continued refinement and systematic comparisons with established frameworks such as EvoSuite~\cite{tang2024chatgpt,bhatia2024unit} and Pynguin~\cite{lemieux2023codamosa,bhatia2024unit}.

\noindent\textbf{Prompt Mutation:} Prompt mutations are effectively utilized in LLM-assisted test and code generation~\cite{yang2024evaluation,lops2024AgoneTest, wang-etal-2023-recode,10.1145/3691620.3695517, chen2025NLPertubator}. Specifically, skeleton prompts enforce generation uniformity~\cite{lops2024AgoneTest} or iterative human-in-the-loop refine requirement~\cite{yang2024evaluation}.  
ReCode~\cite{wang-etal-2023-recode} and ADVPRO~\cite{10.1145/3691620.3695517}, respectively, present robustness-enhancement and paraphrase prompt mutation techniques.
Differing from ReCode and ADVPRO, \tool applies a broader set of sentence-level mutations
and operates entirely on natural-language prompt mutations.
Finally, Chen et al~\cite{chen2025NLPertubator} explore the mutations from the perspective of human-induced variance in semantically similar prompts and
places a significant emphasis upon minor prompt perturbations. 
\tool instead utilizes real-world natural language variance to get semantically similar prompts that can generate diverse code and tests.


\section{Threats to Validity}
\vspace{-3pt}

Firstly, using the same LLM to produce both solutions and tests reduces the configuration effort but risks circularity: if the model generates flawed or overly simplistic tests, incorrect solutions may pass evaluation.
Although our mutation strategies seek to mitigate this by inducing diversity, they cannot guarantee absolute coverage of all edge cases or complex requirements found in real-world software. 
Secondly, our carefully designed diverse set of LLM-based prompt mutations to simulate real-world variation in problem descriptions might not capture the full range or nuance of naturally occurring variances. Additionally, the manually validated sample set indicates GPT-4 has high accuracy; it may still
introduce unintended biases or fail to preserve the original task semantics perfectly.
Finally, while the three evaluation benchmarks, LiveCodeBench, BigCodeBench, and DLBench,  offer complementary coverage—from algorithmic puzzles to real-world API usage and deep learning code—they may not fully represent all software engineering contexts or programming languages.

\section{Conclusion}
\vspace{-3pt}

In this work, we presented \tool, a framework that leverages semantic-preserving prompt mutations and consensus-based verification to enhance the reliability of LLM-driven code generation. 
Evaluation showed that \tool consistently improved pass@1 accuracy when controlled linguistic and structural mutations diversified candidate solutions and strengthened consensus selection. Our analysis revealed complementary benefits across mutation classes
while highlighting trade-offs in token cost when applying larger mutation sets. Overall, \tool provides a practical and extensible approach to improving code generation.

\balance
\bibliographystyle{IEEEtran}
\bibliography{refs}
\end{document}